\documentclass{appolb}
\usepackage{graphicx}
\usepackage[dvipsnames]{xcolor}
\usepackage{slashed}
\usepackage{graphics,bm}
\usepackage{epsfig}
\usepackage{graphicx}
\usepackage{amsmath}
\usepackage{amssymb}
\usepackage{bbold}
\usepackage{wrapfig}
\usepackage{color}
\usepackage{xcolor}
\usepackage{hyperref}
\usepackage{mathtools}
\usepackage{cancel}
\usepackage{multirow}
\usepackage{upgreek}

\begin{document}

\def\a{{\alpha}}
\def\be{{\beta}}
\def\d{{\delta}}
\def\D{{\Delta}}
\def\P{{\Pi}}
\def\p{{\pi}}
\def\e{{\varepsilon}}
\def\ep{{\epsilon}}
\def\g{{\gamma}}
\def\k{{\kappa}}
\def\l{{\lambda}}
\def\L{{\Lambda}}
\def\m{{\mu}}
\def\n{{\nu}}
\def\o{{\omega}}
\def\O{{\Omega}}
\def\S{{\Sigma}}
\def\s{{\sigma}}
\def\t{{\tau}}
\def\x{{\xi}}
\def\X{{\Xi}}
\def\z{{\zeta}}

\def\ol#1{{\overline{#1}}}
\def\c#1{{\mathcal{#1}}}
\def\b#1{{\bm{#1}}}
\def\eqref#1{{(\ref{#1})}}

\def\wt#1{{\widetilde{#1}}}

\def\ed#1{{\textcolor{magenta}{#1}}}
\def\edd#1{{\textcolor{cyan}{#1}}}
\title{Pion Weak Decay in a Magnetic
Field
}
\author{Prabal~Adhikari\thanks{Speaker, presented at the Excited QCD 2026 Workshop.}
\address{Physics Department, 
        Faculty of Natural Sciences and Mathematics,
        St.~Olaf College,
        Northfield, 
        MN 55057, USA}
\\[3mm]
{Brian~C.~Tiburzi 
\address{Department of Physics,
        The City College of New York,
        New York,
        NY 10031, USA}
}
\address{
	Graduate School and University Center,
        The City University of New York,
        New York, 
        NY 10016, 
        USA}
}
\maketitle
\begin{abstract}
Pion decay width in a uniform magnetic background, constructed within chiral perturbation theory, is compared with lattice QCD for which results are available in the muon channel. While the results are consistent for large magnetic fields, the discrepancy observed for weak magnetic fields is largely due to differences in the pion decay constants.
\end{abstract}
  
\section{Introduction}
In a pioneering lattice QCD study~\cite{Bali:2018sey}, a novel pion-to-vacuum matrix element involving a vector current was identified.  The resulting novel pion decay constant $F^{(V)}_{\p}$ was utilized in addition to the usual axial pion decay constants, $F^{(A1)}$ to construct the pion decay width in the muon channel, $\pi^{+}\rightarrow \mu^{+} \nu_{\mu}$. The decay width was constructed under the assumption that the anti-muon in the final decay product decays into the lowest Landau level (LLL). The LLL assumption was relaxed and the full decay width constructed in studies~\cite{Coppola:2018ygv,Coppola:2019wvh,Coppola:2019idh} conducted within the context of the Nambu-Jona-Lasinio (NJL) model. The studies construct the decay width in both the Landau and symmetric gauges and explicitly exhibit gauge independence. Furthermore, the modification of the branching ratio and the angular distribution of the outgoing neutrinos are also investigated.

Chiral perturbation theory, the low-energy effective theory of QCD, provides a model-independent tool to study the pion decay width~\cite{adhikari2024chiral} -- this is unlike the NJL model for which results are model-dependent even when model-insensitive~\cite{adhikari2024chiralsymmetry}. Observables are particularly strongly constrained for weak magnetic fields, $eB\ll 4\p F_{\p}$, a regime in which finite volume corrections are expected to be substantial~\cite{Adhikari:2023fdl} and the LLL approximation is anticipated to fail. For example, the coincident Green's function is not translationally invariant at finite magnetic fields due to Wilson lines -- consequently the chiral condensate exhibits spatial inhomogeneity that depends on the flux quanta. For the lowest flux quanta, the size of finite volume correction in the spatially averaged chiral condensate, for $m_{\p}L=3$, is approximately $25\%$ while for $m_{\p}L=4.5$ the correction is larger than $5\%$. The acuteness of finite volume corrections is anticipated as the order at which finite volume corrections enters chiral perturbation theory is also the order at which magnetic field dependence enters. 

In this letter, we construct, within the model-independent setting of chiral perturbation theory, the pion decay width in the charged sector, $\Gamma_{\pi^{+}\rightarrow\, \ell^{+} \nu_{\ell}}$ and compare the result in the muon channel to that from the seminal lattice investigation.

\section{Effective Lagrangian for Pion Weak Decay}
The effective Lagrangian to investigate pion weak decay follows from chiral perturbation theory~\cite{Gasser:1983yg} and electroweak theory. In \textit{Minkowski} space, the effective Lagrangian, $\c{L}_{\rm eff}=\c{L}_{\p}+\c{L}_{\ell \n}+\c{L}_{\p\ell\n}$,
consists of a purely quadratic Lagrangian associated with pion fields renormalized by the magnetic background, $\c{L}_{\p}$, a quadratic contribution for leptons in the decay product, $\c{L}_{\ell\n}$, and a contact interaction that couples pions to the electroweak sector, $\c{L}_{W\p}$.  The quadratic Lagrangian for magnetized pions, $\c{L}_{\p}$, is 
\begin{align}
\c{L}_{\p}&=D_{\m}\pi^{+}D^{\m}\p^{-}-m_{\p^{\pm}}^{2}(B)\pi^{+}\pi^{-}+\tfrac{1}{2}\partial_{\m}\pi^{0}\partial^{\m}\p^{0}-\tfrac{1}{2}m_{\p^{0}}^{2}(B)\pi^{0}\pi^{0}
\end{align}
with $\p^{\pm}$ denoting renormalized pion fields -- we work to $\c{O}(p^{4})$ in chiral perturbation theory. The covariant derivative of the charged pion fields is $D_{\m}\p^{\pm}=(\partial_{\m}\pm ieA_{\m})\p^{\pm}$ account for Landau levels that are unaccounted for by the renormalized masses, $m_{\p^{\pm}}(B)$. The renormalized charged pion mass receives a perturbative correction~\cite{Tiburzi:2008ma,Andersen:2012dz,Andersen:2012zc,Hofmann:2020ism}
\begin{align}
\label{eq:chargedpionmass}
m^{2}_{\p^{\pm}}(B)&=m^{2}_{\p}\left[1+\frac{eB}{m_{\p}^{2}}\frac{eB}{(4\p F_{\p})^{2}}\ol{l}\right]&
\ol{l}&\equiv\ol{l}_{6}-\ol{l}_{5}=3.0\pm0.3~\cite{Bijnens:2014lea}\ ,
\end{align}
that scales quadratically with the external field at next-to-leading order (NLO) and linearly on the difference of low energy constants, that enters through a scale-independent contribution, $\ol{l}=\tfrac{1}{3}(\ol{l}_{6}-\ol{l}_{5})$. $m_{\pi}$ is the zero-field, degenerate pion mass. The neutral pion interacts with the magnetic background indirectly through tadpole diagrams involving charged pions. The renormalized neutral pion mass
\begin{align}
m^{2}_{\p^{0}}(B)&=m^{2}_{\p}\left[1+\frac{eB}{(4\p F_{\p})^{2}}\c{I}(\tfrac{m_{\p}^{2}}{eB})\right]\ ,
\end{align}
receives perturbative corrections proportional to the amputated charged pion loop or equivalently the coincident charged Green's function~\cite{Adhikari:2023fdl}. Since $\c{I}$ is a negative definite, monotonically decreasing function,
the neutral pion mass decreases upon the introduction of a magnetic field. However, it remains a pseudo-Goldstone boson unlike the charged pion -- in the chiral limit only the neutral pion remains massless. 

The quadratic electroweak Lagrangian, assuming that the $W^{\pm}$ and $Z^{0}$ bosons of the electroweak theory have been integrated out, is
\begin{align}
\c{L}_{\ell\n}&=\bar{\ell}\left(i\slashed{D}-m_{\ell}\right)\ell+\bar{\n}_{\ell}\left(i\slashed{\partial}\right)\n_{\ell}\ .
\end{align}
The charged lepton, $\ell$, couples to the magnetic background through the covariant derivative $\slashed{D}\equiv\g^{\m}(\partial_{\m}+ieA_{\m})$. The (left-handed) neutrino, $\n$, will be assumed to be massless. Systematic corrections associated with utilizing the Lagrangian can be na\"{i}vely estimated through a large mass expansion of the gauge boson propagators and are of relative order ${k^{2}}/{M_{W,Z}^{2}}$. $\c{L}_{\p\ell\n}$ encodes the coupling of the charged pions to the weak sector through the left current, $J_{L}^{\ \m}$ 
\begin{align}
\c{L}_{\p\ell\n}&=J_{W\m}J_{L}^{\ \mu}&
J_{W\m}&={2{G}_{F}V_{ud}\left[\,\ol{\n}_{\ell}\g_{\m}(\mathbb{1}-\g_{5})\ell+\ol{\ell}\g_{\m}(\mathbb{1}-\g_{5})\n_{\ell}\,\right]}\,.
\end{align}
The Fermi coupling, $G_{F}$, depends on the weak coupling constant $g_{W}$ and is associated with the coupling of leptons to the $W^{\pm}$ boson, $G_{F}=\tfrac{1}{4\sqrt{2}}\tfrac{g_{W}^{2}}{M_{W}^{2}}$.
$V_{ud}$  is the first element of the Cabibbo-Kobayashi-Maskawa quark-mixing matrix that rotates weak eigenstates to mass eigenstates of the strong sector. The $W$-boson mass eigenstates and masses are themselves modified by the external fields~\cite{Chernodub:2012fi}. Since the change in the $W$-boson mass is of relative order $\frac{eB}{M_{W}^{2}}$, we ignore this effect. This is justified in the regime of validity of chiral perturbation theory for which the magnetic field is small compared to the typical hadronic scale, $\L_{\chi}\sim 4\p F_{\pi}$, which in turn is significantly smaller than the mass of the $W$-boson, $\sqrt{eB}\ll  4\p F_{\pi} \ll\hspace{-4pt}\ll M_{W}$.
\subsection{Charged Pion Decay Width and Branching Ratio}
\begin{figure}[htb]
\centerline{%
\includegraphics[width=7.cm]{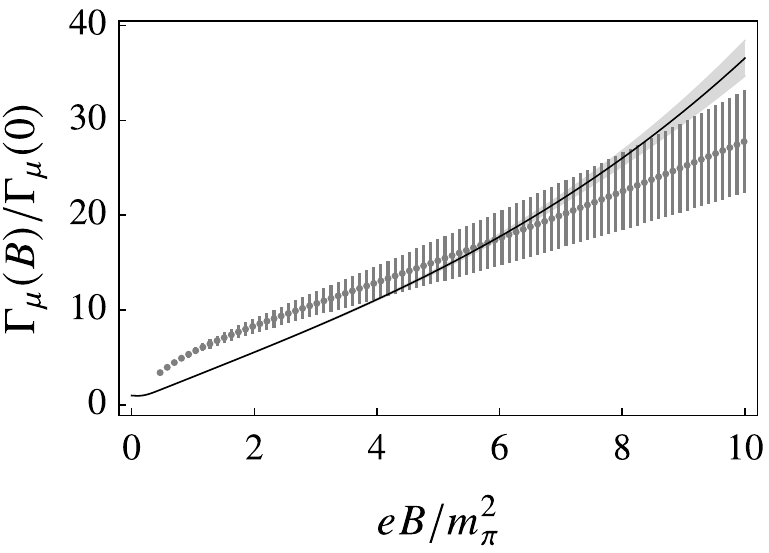}}
\caption{Plot of pion decay width in the $\mu^{+}$ channel for $m_{\pi}=135\ \mathtt{MeV}$ and $\sqrt{2}F_{\pi}=131\ \mathtt{MeV}$. The uncertainty band is produced by varying the uncertainties in low energy constants and varying next-to-next-leading corrections in the range $\pm 2(eB)^{2}/(4\p F_{\p})^{2}$. Corresponding lattice data from Ref.~\cite{Bali:2018sey} is also presented with uncertainties.}  
\label{fig:decay-width-muon}
\end{figure}

At zero magnetic field, the pion decay constant is defined by the pion-to-vacuum matrix element that is proportional to the usual pion decay constant, labelled $F_{\pi}^{(A1)}$ below,
\begin{align}
\langle\,0\,|\, \hat{J}_{A\mu}(x)\,|\,\pi^{Q}(\,P\,)\,\rangle_{x=0}\sim iF_{\pi^{Q}}^{(A1)}\,P_{\mu}\,.
\end{align}
The left-current is supplied by the axial current that comes with an odd-number of pions.  Upon turning on the magnetic field, there are further pion decay constants as there are Lorentz structures available not only through good momentum, $\wt{P}_{\mu}$ but also the magnetic field $F_{\mu\nu}$ and its dual field, $\wt{F}_{\mu\nu}$. Heuristically, the matrix element is
\begin{align} 
\nonumber
&\langle\,0\,|\, \hat{J}_{A\mu}(0)\,|\,\pi^{Q}(\,\widetilde{P}\,)\,\rangle\\
\hspace{-10mm}
\sim&\ i\left[F^{(A1)}_{\pi^{Q}}(B)\,\widetilde{P}_{\mu}+F^{(A2)}_{\pi^{Q}}(B)\,eQF_{\mu\nu}\,\widetilde{P}^{\,\nu}+F^{(A3)}_{\pi^{Q}}(B)\,eF_{\mu\nu}\,eF^{\nu\alpha}\,\widetilde{P}_{\,\alpha}\cdots\right]\,.
\end{align}
The axial current is parity even while the single pion is parity odd consistent with the right hand side where good momentum, $\widetilde{P}$, is parity odd while the magnetic field, $F_{\mu\nu}$ is parity even. For an $\c{O}(p^{4})$ analysis, we only require the contributions of $F^{(A1)}_{\pi^{Q}}$ and $F^{(A2)}_{\pi^{Q}}$ as $F^{(A3)}_{\pi^{Q}}$ first appears at NNLO
\begin{align}
F^{(A1)}_{\pi^{Q}}(B)&=F_{\pi}\left[1-\left(1-\frac{|Q|}{2}\right)\frac{eB}{(4\pi F_{\pi})^{2}}\mathcal{I}(\tfrac{m_{\pi}^{2}}{eB})+\mathcal{O}\left(\frac{eB}{(4\pi F_{\pi})^{2}}\right)\right]\\
F^{(A2)}_{\pi^{Q}}(B)&=\frac{F_{\pi}}{(4\pi F_{\pi})^{2}}\overline{l}\left[|Q|+\mathcal{O}\left(\frac{eB}{(4\pi F_{\pi})^{2}}\right)\right]\,.
\end{align}
There is a further channel involving a left-current that becomes available in a magnetic field due to the Wess-Zumino-Witten (WZW) Lagrangian~\cite{Kaiser:2000ck}. Heuristically, the matrix element is
\begin{align}
\langle\,0\,|\, \hat{J}_{V\mu}(0)\,|\,\pi^{Q}(\,\widetilde{P}\,)\,\rangle&\sim i\left[F^{(V)}_{\pi^{Q}}(B)e\widetilde{F}_{\mu\nu}\widetilde{P}^{\nu}+\cdots\right]\,.
\end{align}
The left-hand-side is parity even since both the vector current and single pion state are parity odd. The dual field $\wt{F}_{\mu\nu}$ is parity odd, which ensures the right-hand-side is also parity even. The resulting vector pion decay constant

\begin{align}
F^{(V)}_{\pi^{Q}}(B)&=\frac{1}{8\pi^{2}F_{\pi}}\left[1+\mathcal{O}\left(\frac{eB}{(4\pi F_{\pi})^{2}}\right)\right]
\end{align}
is readily constructed from the WZW Lagrangian~\cite{Kaiser:2000ck}. 
\begin{figure}[htb]
\centerline{%
\includegraphics[width=7.cm]{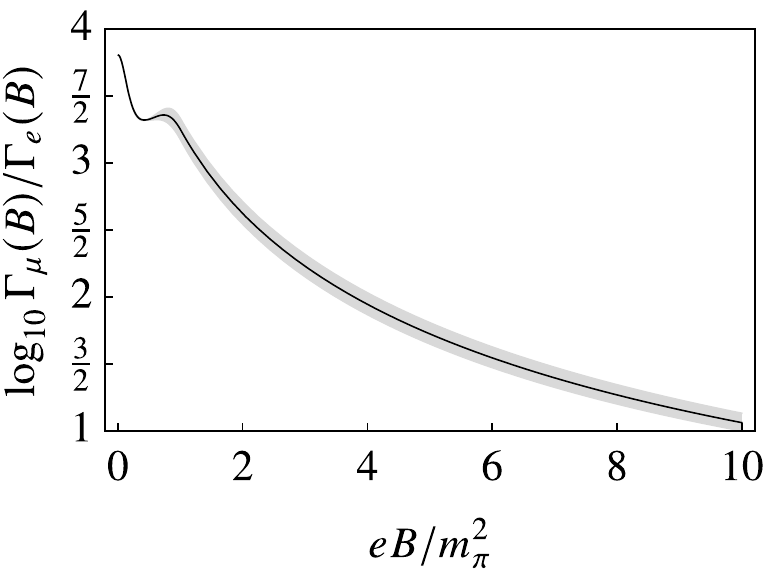}}
\caption{Plot of the branching ratio (for the muon and electron channels) as a function of the magnetic field.}  
\label{fig:decay-width-branching-ratio}
\end{figure}

The resulting pion decay width, $\Gamma_\ell(\,B,P_{z}\,)$,
naturally conserves good momentum and admits a complicated structure unlike at zero field~\cite{adhikari2024chiral}. Under boosts in the $z$-direction, the ratio $\frac{\Gamma_{\ell}(\,B,\ P_{z}\,)\,/\,\Gamma_{\ell}(\,B,\,P_{z}=0)}{E_{0,0}\,/\,E_{0,P_{z}}}$ is one. $E_{0,P_{z}}$ is the pion energy in the lowest Landau level.
The total decay width in the zero field limit is dominated by the muon channel, see Figs.~\ref{fig:decay-width-muon} and \ref{fig:decay-width-branching-ratio}. However, at finite $B$, the decay width in the electron channel increases more rapidly due to the larger (relative) increase in the mass of the electron, the lighter lepton, as opposed to the muon channel, see Fig.~\ref{fig:decay-width-muon}. Consequently, the branching ratio, presented in Fig.~\ref{fig:decay-width-branching-ratio}, decreases from $10^{4}$ to approximately $10$ for $eB=10m_{\p}^{2}$. 

The error bands in the plots indicate uncertainties arising due to the LECs and NNLO contributions. For weak fields, the results are most strongly constrained. Noting that $\lfloor n_{\rm max}\rfloor=1$ for $eB\gtrapprox {0.130}\,m_{\pi}^{2}$ in the muon channel and $eB\gtrapprox {0.334}\,m_{\pi}^{2}$ in the electron channel suggests that the difference between lattice QCD and chiral perturbation theory is not due to the LLL approximation but the pion decay constants. Currently, $F^{(A2)}$ is only available in the latter while there is modest tension with lattice QCD in the zero field limit of $F_{\p}^{(V)}$. However, the qualitative behavior, near $B=0$, of $F^{(A1)}$ extracted through the lattice is inconsistent with the model-independent analysis of chiral perturbation theory~\cite{adhikari2024chiral,adhikari2024chiralsymmetry}.
\bibliographystyle{IEEEtran}
\bibliography{bibly}

\end{document}